\documentclass{article}

\usepackage{arxiv}

\usepackage[utf8]{inputenc} 
\usepackage[T1]{fontenc}    
\usepackage{hyperref}       
\usepackage{url}            
\usepackage{booktabs}       
\usepackage{amsfonts}       
\usepackage{nicefrac}       
\usepackage{microtype}      
\usepackage{lipsum}		
\usepackage{graphicx}
\usepackage[numbers]{natbib}
\usepackage{doi}
\usepackage{tabularx}
\usepackage{comment}

\title{Integration Adapter Architecture for\\Food Traceability Blockchain}

\author{
    \href{https://orcid.org/0009-0005-4862-7800}
    {\includegraphics[scale=0.06]{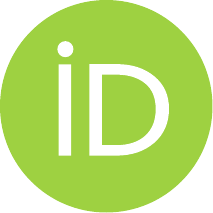}
    \hspace{1mm} André Romão}\\
	INESC-ID, Instituto Superior Técnico,\\Universidade de Lisboa, Portugal.\\
	\texttt{andre.romao@ulisboa.pt} \\
    \And
    \href{https://orcid.org/0000-0001-6350-5506}
    {\includegraphics[scale=0.06]{orcid.pdf}
    \hspace{1mm} Francisco Faria} \\
	INESC-ID, Instituto Superior Técnico,\\Universidade de Lisboa, Portugal.\\
	\texttt{franciscofaria00@ulisboa.pt} \\
    \And
    \href{https://orcid.org/0009-0002-5233-8460}
    {\includegraphics[scale=0.06]{orcid.pdf}
    \hspace{1mm} João R. Matos} \\
    INESC-ID, Instituto Superior Técnico,\\Universidade de Lisboa, Portugal.\\
	\texttt{joao.r.matos@ulisboa.pt} \\
	\And
    \href{https://orcid.org/0009-0003-2081-7712}
    {\includegraphics[scale=0.06]{orcid.pdf}
    \hspace{1mm} Emanuel Nunes} \\
    Sensefinity.\\
    Lisboa, Portugal.\\
	\texttt{emanuel.nunes@sensefinity.com} \\
    \And
	\href{https://orcid.org/0000-0003-0972-4171}
    {\includegraphics[scale=0.06]{orcid.pdf}
    \hspace{1mm} Samih Eisa} \\
    INESC-ID, Instituto Superior Técnico,\\Universidade de Lisboa, Portugal.\\
	\texttt{samih.eisa@inesc-id.pt} \\
    \And
	\href{https://orcid.org/0000-0003-2872-7300}
    {\includegraphics[scale=0.06]{orcid.pdf}
    \hspace{1mm} Miguel L. Pardal} \\
    INESC-ID, Instituto Superior Técnico,\\Universidade de Lisboa, Portugal.\\
	\texttt{miguel.pardal@tecnico.ulisboa.pt} \\
}

\renewcommand{\shorttitle}{FoodSteps Integration Adapters}

\hypersetup{
pdftitle={Integration Adapter Architecture for Food Traceability Blockchain},
pdfsubject={Enterprise Integration and Blockchain Systems},
pdfauthor={André Romão, Francisco Faria, João R. Matos, Emanuel Nunes, Samih Eisa, Miguel L. Pardal},
pdfkeywords={Enterprise Integration, Integration Adapters, Blockchain, Hyperledger Fabric, Food Traceability, Message Queues},
}

\begin{document}
\maketitle

\begin{abstract}
Enterprise adoption of permissioned blockchains remains limited due to the complexity and cost of integrating legacy systems. We present a modular adapter architecture that bridges enterprise applications with blockchain networks, designed to support small and medium-sized enterprises with limited technical resources. The architecture provides five key modules: (1) configurable data extractors supporting diverse interfaces such as APIs and file uploads, (2) data transformers that can convert to standard formats, (3) messaging middleware to ensure operations can tolerate lack of connectivity and traffic spikes, (4) blockchain loader to commit transactions to the blockchain, and
(5) status visibility to collect and expose runtime metrics that support operational transparency.
We validated the adapters through a pilot deployment in a real-world fruit supply chain, involving three distinct enterprises. The pilot achieved blockchain integration with minimal workflow disruption, demonstrating the usefulness of these adapters for practical interoperability of existing systems with the blockchain.  
\end{abstract}

\keywords{Enterprise Integration, Integration Adapters, Blockchain, Hyperledger Fabric, Message Queues.}

\setcounter{footnote}{0}

\section{Introduction}

Over the past decade, permissioned blockchains have been promoted as a means to improve accountability and trust in enterprise environments. By providing tamper-evident data sharing among authorized participants, they promise significant benefits in sectors such as finance~\cite{tapscott2016blockchain}, healthcare~\cite{Angraal2017BlockchainTA}, and supply chains~\cite{Kshetri2017BlockchainsRI,Tian2016AnAS}. Despite these potential advantages, enterprise adoption has remained limited. A key barrier is the difficulty and cost of integrating legacy systems with blockchain networks, formed by multiple trading partners~\cite{Pazaitis2017BlockchainAV}. In particular, many small and medium-sized enterprises (SMEs) lack the technical resources required 
to participate in these blockchain-based systems.

Existing integration approaches typically rely on custom connectors or tightly coupled middleware~\cite{Olech2024FromTT}. These solutions tend to be costly, inflexible, and disruptive to established workflows. Furthermore, they often fail to account for the resilience and scalability needs of production environments, such as tolerating intermittent connectivity or handling traffic spikes. As a result, the potential of permissioned blockchains to enable trustworthy and collaborative enterprise processes remains under-realized~\cite{Androulaki2018HyperledgerFA}.

In this paper, we propose a modular adapters architecture designed to simplify and standardize the integration of enterprise applications with permissioned blockchains. The architecture consists of configurable data extractors, flexible data transformers, messaging middleware, and dedicated modules for blockchain loading and status visibility. This modular design allows organizations to adopt blockchain technology with minimal workflow disruption while still ensuring extensibility, security, and operational transparency.

We validate the proposed architecture through a pilot deployment in a food traceability blockchain.
In particular, the considered blockchain is dedicated to the traceability of Fundão Cherries, from farm to fork. This scenario provided a testbed involving heterogeneous systems across farms, factories, and retailers. The deployment demonstrated that the proposed adapters architecture could integrate diverse enterprise systems efficiently, without imposing significant technical burdens on participants.

The main contributions of this paper are:
\begin{itemize}
    \item A modular adapter architecture that supports integration of enterprise systems with permissioned blockchains;
    \item A pilot deployment in a real-world supply chain that validates the architecture’s effectiveness in practice;
    \item An evaluation combining functional testing and performance benchmarking.
\end{itemize}

The remainder of the paper is organized as follows. Section 2 provides background topics. Section 3 reviews related work. Section 4 presents the adapters architecture. Section 5 describes the cherry supply chain use case. Section 6 describes the implementation of the adapters, while Section 7 reports on the deployment and evaluation. Section 8 discusses key lessons and limitations, and Section 9 concludes the paper.

\section{Background}
This section gives an overview of background topics related to enterprise application integration, core concepts of permissioned blockchains, and challenges associated with integrating these technologies.

\subsection{Enterprise Application Integration}
Enterprises operate a variety of heterogeneous systems, ranging from legacy databases to modern ERP and API-based services. Integration has long been a concern, addressed by Enterprise Application Integration (EAI) solutions such as message-oriented middleware and service-oriented architectures~\cite{Fruehauf2016EnterpriseIP}. In supply chains, standards like EPCIS provide a shared data model for event capture and exchange~\cite{epcis2014}. While these solutions facilitate interoperability, they are typically limited to intra-organizational contexts and do not address cross-enterprise trust and governance.

\subsection{Permissioned Blockchains in Enterprise Settings}

Blockchains provide a shared, tamper-evident ledger for multi-party collaboration. Public blockchains, however, raise concerns about scalability, confidentiality, and governance. Permissioned blockchains, such as Hyperledger Fabric~\cite{Androulaki2018HyperledgerFA} and Quorum~\cite{quorum2016}, overcome these by enforcing membership controls, fine-grained access policies, and configurable consensus. They are increasingly used in domains requiring traceability, provenance, and auditability, including agri-food~\cite{Tian2016AnAS}, pharmaceuticals~\cite{Haq2018BlockchainTI}, and manufacturing~\cite{Chaikaew2022BlockchainFS}. By enabling immutability and distributed trust, permissioned blockchains form a strong foundation for enterprise data integration.


Despite these benefits, adoption remains limited. Connecting enterprise applications with blockchains introduces key challenges:
    a) system heterogeneity where data originates from diverse formats and interfaces (e.g., files, APIs, databases);
    b) workflow disruption indicates that SMEs often lack resources for costly IT transformations;
    c) scalability and resilience where integration pipelines must handle workload bursts and tolerate intermittent connectivity;
    d) security to ensure strong authentication, authorization, and validation must be enforced across organizational boundaries~\cite{Xu2019ArchitectureFB}.


These challenges highlight the need for lightweight, modular integration mechanisms that support security and resilience while minimizing technical overhead.

\section{Related Work}
\label{sec:relatedWork}

Blockchain-based solutions have been widely explored to improve transparency, provenance, and trust in supply chains. However, practical adoption, especially by small and medium-sized enterprises (SMEs), remains challenging due to technological complexity, integration costs, and governance constraints \cite{Pearson2019}.

\subsection{Enterprise Blockchain Platforms}

To reduce entry barriers, several enterprise-oriented blockchain platforms have emerged, often delivered as Blockchain-as-a-Service (BaaS). Platforms such as Kaleido and Azure Blockchain Service abstract infrastructure management by providing managed networks, pre-configured nodes, and SDK-based access \cite{singh2018blockchain}. While these offerings simplify deployment, they primarily focus on infrastructure provisioning and leave data integration and semantic interoperability largely unaddressed.

Building on BaaS, sector-specific platforms have been developed for supply chain traceability. IBM Food Trust \cite{kamath2018food} is a prominent example in the food domain, offering predefined data models (e.g., EPCIS events), standardized APIs, and consortium-based governance. Its deployment by Walmart demonstrated substantial reductions in traceability time \cite{IBMWalmart}. However, participants do not operate blockchain nodes and must conform to platform-defined schemas and workflows, limiting flexibility.

TradeLens \cite{Jovanovic2022}, developed by IBM and Maersk, applied a similar model to maritime logistics. Despite robust technical foundations, it was discontinued in 2023 due to insufficient industry adoption, high participation costs, and concerns over data governance. Everledger \cite{Smits2020}, targeting high-value assets such as diamonds, focuses on provenance and authenticity but operates as a largely provider-controlled system with limited transparency regarding infrastructure governance.

Overall, these enterprise platforms reduce technical barriers but rely on centralized or consortium-controlled architectures, often constraining customization, interoperability, and equitable value distribution among participants. Table~\ref{tab:comparison} presents a comparison of these approaches.

\subsection{Blockchain Gateways and Integration Approaches}

An alternative line of work focuses on blockchain gateways and middleware. Gateways act as intermediaries for transaction submission, identity management, and access control \cite{Hardjono2018TowardsAD}. While they simplify interaction with distributed ledgers, they frequently become tightly coupled to specific blockchain implementations and risk introducing single points of failure.

Numerous research prototypes explore blockchain integration with supply chains and IoT systems. Tian \cite{Tian2016AnAS} proposed a blockchain-based traceability system for agri-food products, and Reyna et al.~\cite{reyna2018iot} surveyed blockchain–IoT integration architectures. These works demonstrate the feasibility of blockchain-based provenance tracking but typically rely on custom-built connectors and domain-specific assumptions, limiting reuse across contexts.

Generic enterprise integration frameworks, such as Apache Camel and MuleSoft \cite{Fruehauf2016EnterpriseIP}, provide powerful mechanisms for system interoperability but are not tailored to blockchain-specific concerns such as immutability, provenance semantics, or ledger governance. Blockchain-focused middleware frameworks, including Hyperledger Cactus \cite{cactus2020} and Weaver \cite{weaver2021}, aim to support interoperability across heterogeneous ledgers. Although flexible, these solutions demand significant configuration effort and technical expertise, which can be prohibitive for SMEs.

\begin{table}[t]
\centering
\caption{Comparison of blockchain-based traceability approaches}
\scriptsize
\label{tab:comparison}
\begin{tabular}{lcccc}
\hline
\textbf{Feature} & \textbf{IBM Food Trust} & \textbf{TradeLens} & \textbf{Everledger} & \textbf{Proposed Approach} \\
\hline
Domain focus & Food supply chain & Maritime logistics & High-value assets & Food supply chains \\
Blockchain type & Permissioned & Permissioned & Permissioned & Permissioned \\
Infrastructure control & Centralized (IBM) & Centralized (IBM) & Centralized & Decoupled via adapters \\
Governance model & Consortium-led & Consortium-led & Provider-led & Organization-controlled \\
Node operation by users & No & No & No & Optional / abstracted \\
Data integration & Platform-specific APIs & Platform-specific APIs & REST APIs & Adapter-based mediation \\
Schema flexibility & Limited & Limited & Limited & High (canonical + adapters) \\
Support for heterogeneity & Moderate & Moderate & Moderate & High \\
SME accessibility & Medium & Low--Medium & Medium & High \\
\hline
\end{tabular}
\end{table}

\section{Adapters Design}
This section describes the adapters' design, whose purpose is to seamlessly bridge heterogeneous enterprise applications with permissioned blockchain networks. The architecture is guided by three key principles:
\begin{itemize}
    \item Modularity: each function is encapsulated as an independent component that can be replaced or extended without affecting the others.
    \item Technology Agnostic: adapters operate across diverse enterprise systems, from RESTful APIs to legacy file-based exports.
    \item Resilience and Scalability: built-in buffering and asynchronous messaging handle intermittent network outages and traffic spikes.
\end{itemize}

\begin{figure}
    \centering
    \includegraphics[width=0.9\columnwidth]{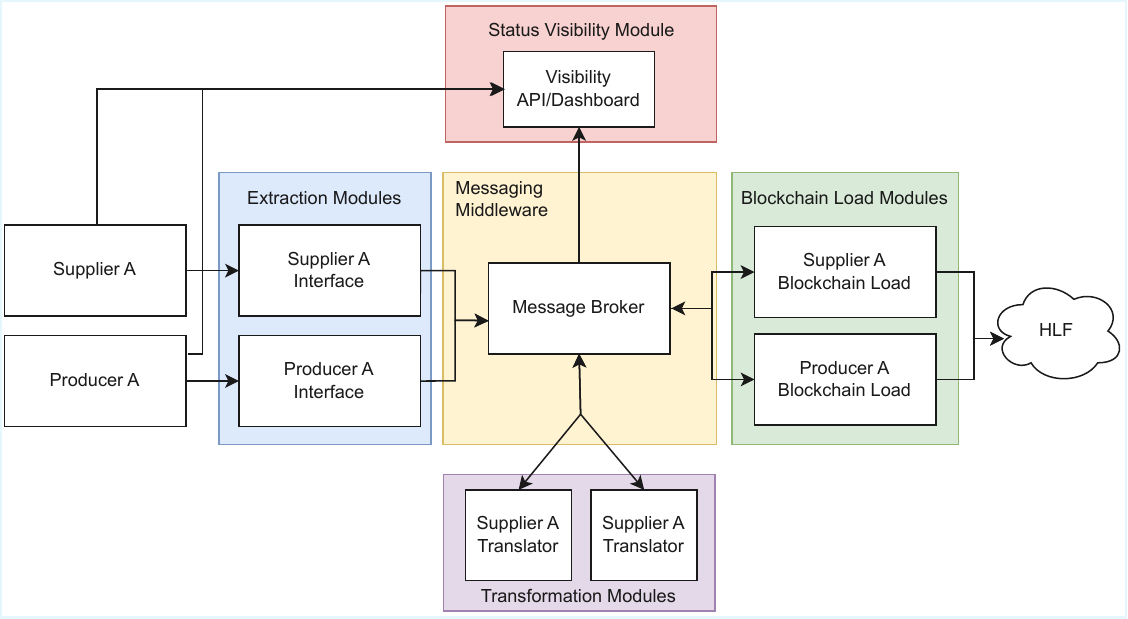}
    \caption{High-level architecture of the solution.}
    \label{fig:highLevelOverview}
\end{figure}

Figure~\ref{fig:highLevelOverview} depicts the overall architecture. The adapters sit between enterprise systems and the target blockchain network. Data flows from enterprise sources (i.e., supplier and producer) through Extractors, passes through a Transformer Layer for schema harmonization, and then published to the blockchain via a Messaging Middleware. Configuration files define each stage so that enterprises can adapt the pipeline without code changes. The architecture follows a layered and modular design that separates data acquisition, transformation, and blockchain interaction concerns. This separation ensures that updates or failures in one module do not affect the operation of others, improving maintainability and scalability. The architecture, as represented in Figure~\ref{fig:highLevelOverview}, decomposes the integration layer into five main modules:

\subsection{Extraction Modules}
These connectors interface with a variety of enterprise data sources, including REST APIs, message queues, and scheduled file uploads (e.g., CSV or JSON). They support both active data retrieval (e.g., database polling, API calls) and passive reception (e.g., file uploads), enabling integration with partners of varying technical capabilities.
Each extractor is defined by a YAML configuration describing endpoint credentials, polling intervals, and data format.

\subsection{Transformation Modules}
The transformers map incoming data to a canonical schema suitable for blockchain transactions. They support both declarative transformations (e.g., JSON-to-JSON mapping) and custom transformation scripts for complex logic, ensuring consistency across heterogeneous systems. It verifies event structure against predefined schemas, rejecting malformed inputs and executing domain-specific conversions to ensure interoperability. This staged architecture isolates validation and mapping from protocol and blockchain logic.

\subsection{Messaging Middleware}
The messaging middleware layer ensures reliable, ordered delivery of transformed records to the blockchain network. 
It implements durable queues and back-pressure control to tolerate temporary network failures or sudden surges in transaction volume. 
Logical isolation routes messages to dedicated transformation pipelines based on origin identifiers, preserving data sovereignty.
Middleware instances can be deployed redundantly for high availability.

\subsection{Blockchain Loader Modules}
The blockchain loader modules encapsulate the logic for committing standardized transactions to the underlying permissioned blockchain.
It consumes transformed messages from the middleware and manages the complete Fabric transaction workflow.
Each event triggers a chaincode invocation on the company’s private channel, using the event’s cryptographic hash as an idempotency key to prevent duplicates.

\subsection{Status Visibility Modules}

The Status Visibility Module provides end-to-end observability across all integration stages, addressing a limitation of native blockchain tooling. Each request transitions through five states: i) Received, upon successful ingestion; ii) Translated, after EPCIS validation and mapping; iii) Processing, during blockchain endorsement and ordering; iv) Confirmed, once included in a block; v) Failed, if validation or persistence fails.

State transitions are persisted in an audit log and correlated with Fabric transaction identifiers and block metadata via Fabric’s event mechanism. This enables participants to verify delivery guarantees and transaction finality through APIs or dashboards. Beyond operational monitoring, this visibility layer lays the foundation for higher-level traceability services, such as reconstructing product journeys indexed by GS1 identifiers.

\section{Use Case: Cherry Supply Chain Traceability with Adapter Services}
\label{sec:usecase}

\begin{figure*}[h]
    \centerline{\includegraphics[width=0.90\columnwidth]{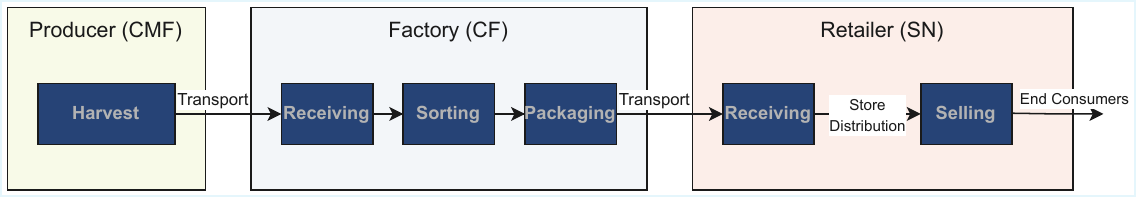}}
    \caption{Cherry flow along the supply chain: from producer, to factory, to retailer, and to consumers.}
    \label{fig:flowPilot}
\end{figure*}

To evaluate the proposed adapter services in a realistic and operational environment, we conducted a pilot deployment within the cherry supply chain in the Fundão region of Portugal. This region is renowned for cherries certified under a Protected Geographical Indication (PGI), a designation that enforces strict requirements on product origin, handling, and quality. 
These regulatory constraints make end-to-end traceability essential and provide a suitable context for validating the integration mechanisms.
The pilot involved three primary organizations spanning the cherry supply chain\footnote{Actual company names are omitted.}, as illustrated in Figure~\ref{fig:flowPilot}:

\begin{itemize}
    \item CMF Farm (Production): responsible for harvesting and initial quality checks;
    \item CF Processing Company (Factory): handles packaging, labeling, and aggregation of harvest batches;
    \item SN Retailer (Distribution and Sales): manages shipments and final retail operations.
\end{itemize}

Each organization operates its own information systems and data sources, including farm management tools, factory logistics systems, and retail inventory platforms. In addition, sensor data and manual records are generated during harvesting, packaging, and transportation. Instead of interfacing directly with the blockchain infrastructure, all provenance-related data are submitted through an Adapter Service that acts as an integration and mediation layer. It validates incoming data against a canonical traceability model, normalizes measurement units and timestamps, enriches records with contextual metadata (e.g., actor identity, location, and process stage), and performs consistency and completeness checks. The transformed data are then packaged into ledger-ready transactions and committed to a permissioned Hyperledger Fabric network, ensuring immutability and auditability.

\begin{figure}[htp]
    \centering
    \includegraphics[width=0.90\textwidth]{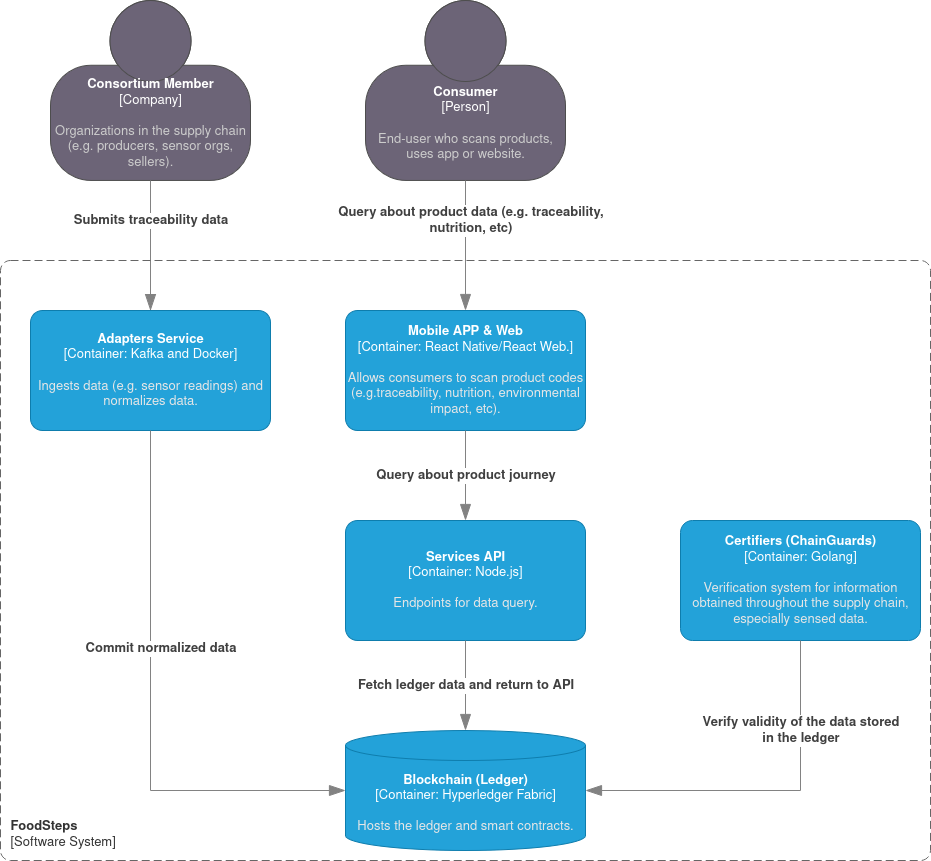}
    \caption{FoodSteps architecture highlighting the role of adapter services}
    \label{fig:foodsteps}
\end{figure}

The pilot of the adapter services that integrate heterogeneous data sources across the supply chain was implemented on FoodSteps, a blockchain-based food traceability platform. Figure~\ref{fig:foodsteps} presents the overall FoodSteps architecture, 
using C4 notation\footnote{Context, Containers, Components, and Code diagrams. \url{https://c4model.com/}}. The figure highlights the role of adapter services as intermediary between data-producing systems and the provenance ledger.

Hyperledger Fabric’s channel mechanism is used to enforce data governance and confidentiality. Sensitive operational data remain confined to private channels controlled by individual organizations, while selected records are published to shared channels when cross-organizational traceability is required, for example after a sale or shipment is completed. This approach enables end-to-end traceability while preserving commercial confidentiality.

\begin{figure}[t]
\centering
\includegraphics[width=0.32\linewidth]{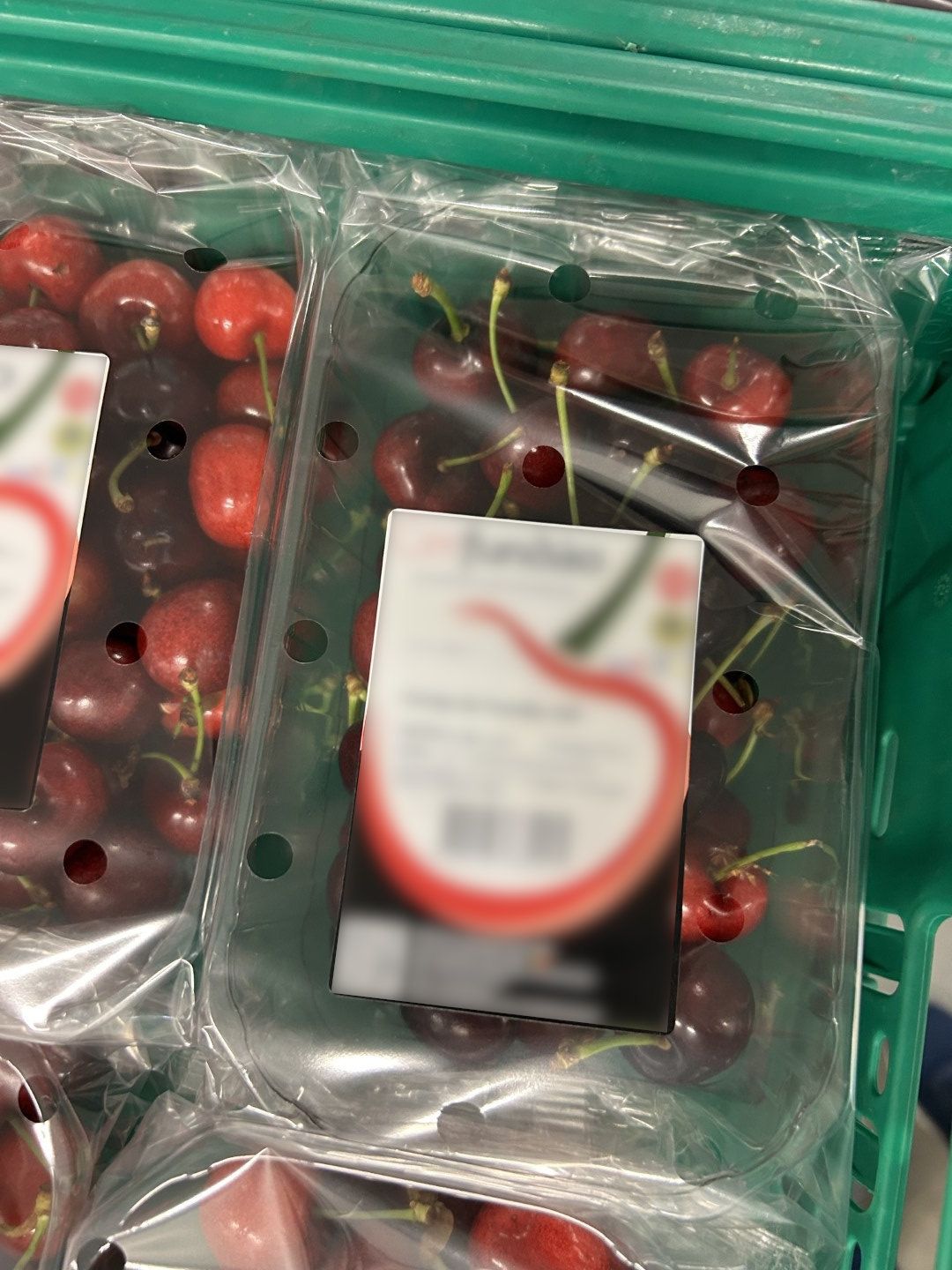}\hspace{0.02\linewidth}
\includegraphics[width=0.32\linewidth]{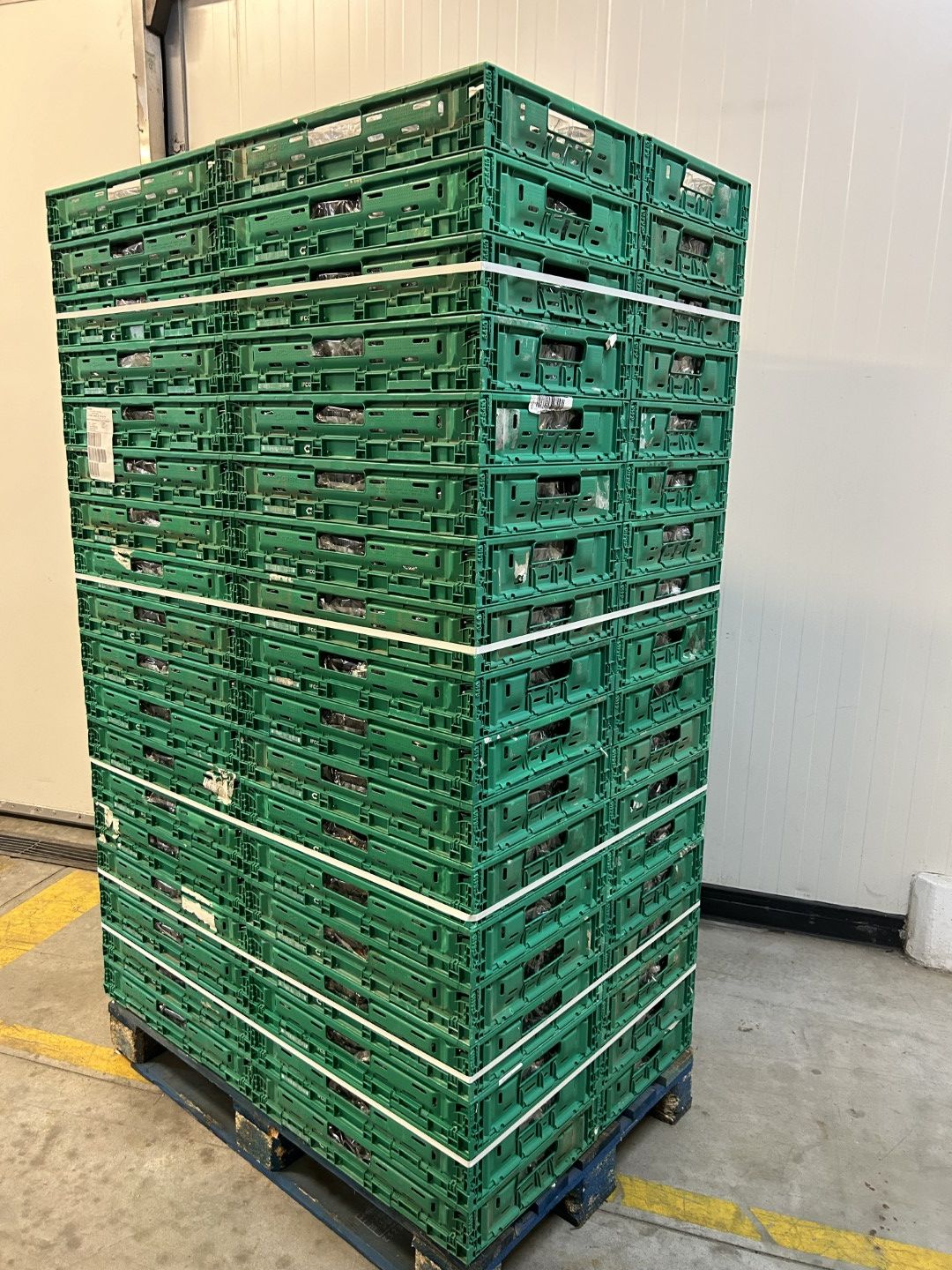}
\caption{Photos at the cherry processing factory:
(left) cherries cuvette with identifier codes;
(right) pallet of cuvettes ready for shipping.}
\label{fig:cherriespallete}
\end{figure}

The pilot study was conducted over approximately one year and included meetings with farm managers, factory logistics staff, and retail managers to analyze existing workflows, data formats, authentication mechanisms, and traceability practices. Multiple on-site visits were carried out at harvesting fields, processing facilities, and retail outlets. Figure~\ref{fig:cherriespallete} shows examples from the processing stage, including labeled cherry cuvettes and pallets prepared for shipping.

Consumers and external stakeholders access traceability information through web or mobile applications by scanning a QR code or entering a product identifier. These applications query a Services API, which retrieves blockchain-backed records that have been processed and validated by adapter services. The resulting view presents a verified and structured history of the product lifecycle, supporting transparency, regulatory compliance, and informed consumption.

This use case demonstrates how adapter services enable scalable and reusable integration across heterogeneous actors and systems, serving as a key architectural component for blockchain-based food traceability solutions.

\section{Adapters Implementation}
\label{sec:implementation}

This section describes the implementation of the adapters across the Fundão cherries supply chain. It details the extraction, transformation, and loading of data from the several entities involved in this process.
Table~\ref{tab:techstack} summarizes the main technologies and frameworks used in the implementation. 

\begin{table}[t]
\centering
\caption{Technology stack used for adapters' implementation.}
\label{tab:techstack}
\scriptsize
\begin{tabular}{p{1.6cm} p{2.2cm} p{3.5cm}}
\toprule
\textbf{Component} & \textbf{Technology / Tool} & \textbf{Purpose} \\
\midrule
Programming & Python, Go & Service orchestration and high-performance routines \\
Blockchain & Hyperledger Fabric v2.5 & Permissioned ledger for traceability and provenance \\
Messaging & Apache Kafka (3 brokers) & Reliable, ordered data delivery between modules \\
Containerization & Docker, Nginx & Service isolation, portability, secure routing \\
Monitoring & Prometheus, Grafana & System visibility and performance tracking \\
Security & TLS 1.3, API keys & End-to-end encryption and access control \\
\bottomrule
\end{tabular}
\end{table}

\subsection{Infrastructure and Security}

To standardize data ingestion and allow interoperability across heterogeneous systems, the consortium agreed on a common event format: EPCIS~\cite{gs1traceability}. Each record stores \emph{Who}, \emph{What}, \emph{When}, \emph{Where}, \emph{Why} and optionally \emph{How} for an event in the supply chain that is relevant for traceability. The adapters were developed and hosted at premises of the 
development entities during the pilot phase. Despite this, the architecture was designed to support future decentralization, allowing each enterprise to host and operate its own adapter and blockchain node. All adapter modules were containerized using Docker and orchestrated within a fault-tolerant Kafka cluster (three brokers). Kafka ensured high availability, message ordering, and persistence. The communications between all components used TLS 1.3 encryption, while the authentication relied on passwords or API keys. A reverse-proxy layer implemented via Nginx handled routing while preserving end-to-end encryption (TLS pass-through mode). Each module was configured with Docker restart policies to enable automatic recovery from failures. The adapters layer employed a fault-tolerant Kafka cluster (three brokers) as the messaging backbone, with company-specific topics to isolate data and filter messages for downstream modules.

\subsection{Farm Integration}

At the CMF farm, data on harvest lots was already recorded in the WI digital farm-management system. To enable integration without modifying this system, a web-scraping–based extractor was implemented. A dedicated user account was registered in the WI system with read-only access to the harvest registry page. Then the extractor module periodically (hourly) scrapes new entries containing identifiers, weight, variety, and harvest date, and publishes them to CMF’s Kafka topic. The transformation module subscribed to this topic converts the data into EPCIS event format, which is then written back to Kafka. Finally, the Blockchain Load Module commits the EPCIS records to CMF’s private channel on the Hyperledger Fabric network.

\subsection{Factory Integration}
At the CF factory, operations were already managed through the OL logistics system. Following discussions with OL’s technical team, a REST-based API adapter was designed. Three endpoints were defined to capture distinct event types in the production workflow: \texttt{/entryBatch} called when entry batch identifiers are created;\texttt{/manageBatches} triggered when exit batches are created, including information about which entry batches were used; \texttt{/exitBatch} sent when exit batches are added to a shipment.

Each endpoint requires username and API-key authentication through request headers. Upon validation, the handler writes incoming payloads to CF’s Kafka topic. The transformation module converts these messages into EPCIS events, and the Blockchain Load Module submits them to CF’s private ledger channel.

\subsection{Retailer Integration}
The SN retailer used an ERP system that tracks product arrivals and shipments but did not provide API access. To accommodate this constraint, a semi-manual adapter was developed. Retail employees generate a daily file with events data and upload it through a secure web interface. The interface authenticates users via credentials and supports drag-and-drop uploads with feedback on validation errors. The transformation module processes the file, ignoring metadata rows and converting each data line into an EPCIS event. These records are then published to SN’s Kafka topic and later committed to its blockchain channel. This low-complexity workflow preserved data security and consistency while matching SN’s operational constraints.

\section{Evaluation}
The evaluation proceeded in three phases:
\begin{enumerate}
    \item Functional Testing: Module-level validation of critical components such as data ingestion, transformation, and blockchain write operations;
    \item Performance Analysis: Measurement of runtime characteristics such as latency, throughput, and error rates, under simulated and real workloads;
    \item Requirements Assessment: Verification that both functional and non-functional requirements were met, ensuring the completeness of the proposed solution.
\end{enumerate}

Monitoring was conducted using a Prometheus exporter\footnote{https://prometheus.io/} (kafka-exporter) and Grafana\footnote{https://grafana.com/} dashboards implemented to track message production, consumption rates, and system resource usage.

\subsection{Functional Testing}
Each adapter module was tested in isolation and within the full integration workflow to ensure end-to-end consistency. 
A comprehensive suite of tests was developed to validate all implemented operations and data handling. 

Authentication and Authorization were systematically verified, data integrity across the full pipeline was proven, starting with Input Validation on extraction and transformation module. 
The EPCIS Transformation modules were validated in two phases: initial unit-testing verified correct standard output generation, followed by real-time environment testing where the module successfully read raw data from Kafka and outputted transformed EPCIS data back to the middleware. 
A locally deployed Kafka-webviewer instance confirmed the traceability and expected format of the generated EPCIS data. 
Furthermore, Kafka message routing correctness was verified, ensuring that producers and consumers exclusively interacted with their designated topics with zero misrouting occurrences. A blockchain writing test was developed to confirm the correct implementation of the duplication detection mechanism, by sending multiple data entries.

Table~\ref{tab:functional-testing} summarizes the outcomes of the functional tests that were performed across all core modules. 

\begin{table*}[h]
\centering
\caption{Summary of functional testing.}
\label{tab:functional-testing}
\small
\begin{tabularx}{\textwidth}{
  >{\raggedright\arraybackslash}p{2.4cm}
  >{\raggedright\arraybackslash}X
  >{\raggedright\arraybackslash}X}
\toprule
\textbf{Function} & \textbf{Test Focus / Description} & \textbf{Outcome} \\
\midrule
Authentication &
API key and username verification using Postman with valid/invalid credentials &
Valid: \texttt{200 OK}; Invalid: \texttt{401 Unauthorized} with JSON error detail \\
\addlinespace
Input Validation &
Correct vs.\ malformed payloads sent to extraction and transformation modules &
Malformed payloads rejected with precise messages (API response or visibility dashboard) \\
\addlinespace
Transformation &
EPCIS conversion tested both directly and via Kafka subscription &
EPCIS events matched standard schema (verified via kafka-webviewer) \\
\addlinespace
Kafka Routing &
Producers/consumers restricted to designated topics; check for leakage &
No cross-topic leakage; all messages correctly routed \\
\addlinespace
Blockchain Writing &
Duplicate detection via identical JSON payloads in different orders &
Duplicates detected and suppressed; single commit only \\
\bottomrule
\end{tabularx}
\end{table*}

\subsection{Performance Analysis}
Load tests were conducted to evaluate the adapters under conditions that are expected to exceed the operational load of the pilot deployment. These experiments assessed the system’s ability to maintain low latency, stable throughput, and reliable message delivery during sustained and peak traffic scenarios. 
Using python's k6\footnote{https://github.com/grafana/k6}, were initially executed with a peak of 100 requests per second and later increased to 500 requests per second. Each test was generating approximatly 10MB of test data.
The key performance results are summarized in Table~\ref{tab:performance-analysis}.

\begin{table*}[!h]
\centering
\caption{Performance analysis.}
\label{tab:performance-analysis}
\small
\begin{tabularx}{\textwidth}{
  >{\raggedright\arraybackslash}p{4.2cm}
  >{\raggedright\arraybackslash}p{3.5cm}
  >{\raggedright\arraybackslash}X}
\toprule
\textbf{Metric} & \textbf{Result} & \textbf{Notes} \\
\midrule
Initial authentication latency & $\sim$800\,ms/req & Due to Go \texttt{bcrypt} hashing (intentionally slow); optimized for pilot \\
Optimized authentication latency & $\sim$20\,ms/req & Faster (less robust) hash used only for pilot testing \\
API latency (100 req/s, no Kafka) & 22\,ms avg & Measured with Python k6-style load; no data loss \\
API latency (100 req/s, with Kafka) & 24\,ms avg & $\approx$8\% overhead from Kafka enqueue/dequeue \\
Sustained load & 500 req/s for 1 min & Producers met target; no data loss observed \\
Kafka consumer throughput & $\approx$2{,}500 msg/min & Acceptable given Fabric $\approx$1 tx/s sequential commit \\
Kafka log retention & 512\,MB per broker & Supports $\sim$1M standard traceability events \\
Broker memory usage & $\sim$1\,GB per broker & Stable under peak load \\
\bottomrule
\end{tabularx}
\end{table*}

\subsection{Requirements Assessment}

Table~\ref{tab:reqAssessment} presents an assessment of the identified requirements. 

\begin{table}[htbp]
\caption{Requirements assessment.}
\label{tab:reqAssessment}
\centering
\small
\begin{tabular}{lllc}
\toprule
\textbf{ID} & \textbf{Requirement} & \textbf{Status} \\
\midrule
R1 & Flexible Data Ingestion & \checkmark \\
R2 & Enforce Data Validation & $\circ$ \\
R3 & Translate to EPCIS & \checkmark \\
R4 & Blockchain Writing & $\circ$ \\
R5 & Exactly Once Delivery & \checkmark \\
R6 & Transaction Visibility & \checkmark \\
R7 & Decoupling & \checkmark \\
R8 & Fault Tolerance & \checkmark \\
R9 & Security & $\circ$ \\
R10 & Extensibility & \checkmark \\
R11 & Modularity & \checkmark \\
\bottomrule
\end{tabular}
\par\vspace{0.2cm}
\footnotesize
\begin{tabular}{p{\linewidth}}
\center{$\checkmark$ = Fully Met, $\circ$ = Partially Met} \\
\end{tabular}
\end{table}

Testing confirmed that most requirements were satisfied. Three items remain partially fulfilled:
\begin{itemize}
    \item R2 - Enforce Data Validation: current validation occurs in the adapters, but chaincode-level checks on EPCIS data are not yet implemented;
    \item R4 – Blockchain Writing Across Multiple Networks: thorough testing is still needed when a single blockchain load module writes to multiple ledgers simultaneously;
    \item R9 – Security: a full independent security audit remains to be completed.
\end{itemize}

These identified gaps will guide future development of the adapters toward stronger end-to-end guarantees.

\section{Discussion}
The modular adapter architecture used in the Fundão cherry supply chain pilot shows how separating data extraction, transformation, messaging, and blockchain loading allows each participant to connect their legacy systems with minimal code changes and without disrupting existing workflows.

The adapter’s plug-and-play design also made it easier for external blockchain service providers to host nodes on behalf of the supply-chain actors, which is essential for SMEs lacking dedicated IT teams. The fault-tolerant Kafka backbone was essential for handling network fluctuations and transaction bursts during peak harvest periods.

While the architecture already provides modularity, communication resilience, and security, we identified areas for future improvement: on-chain validation of EPCIS data, multi-network writing support, and performance improvements. End-to-end data validation remains incomplete, as chaincode-level validation of EPCIS events has yet to be implemented, leaving a potential gap in on-ledger data assurance. The blockchain loader module also requires additional testing for multi-network operation, which is particularly important for enterprises participating in overlapping supply chains. 

Finally, scalability remains a key consideration: performance tests show that Kafka consumers currently handle approximately 2,500 messages per minute, while Hyperledger Fabric’s default sequential commit rate is about one transaction per second. This throughput is sufficient for the pilot deployment, but large-scale adoption will require parallel consumer partitions and possibly a revised consensus configuration to achieve higher transaction volumes.

\section{Conclusion}
This work presented a modular adapter architecture that enables enterprises, especially small and medium-sized (SMEs), to integrate heterogeneous legacy systems with permissioned blockchain networks. The design separates data extraction, transformation, messaging, blockchain loading, and status visibility, allowing each component to be deployed, scaled, and maintained independently.

A real-world pilot in the Fundão Cherry Chain demonstrated the practicality of this approach. Our evaluation showed that the adapters deliver low-latency, lossless data transfer, meet most functional and non-functional requirements, and can be deployed with minimal disruption to existing systems and connect them to a permissioned blockchain supported by a supply chain consortium.

\clearpage 

\section*{Acknowledgements}

Work supported by national funds through Fundação para a Ciência e a Tecnologia, I.P. (FCT) under projects UID/50021/2025 (DOI: \url{https://doi.org/10.54499/UID/50021/2025}) and UID/PRR/50021/2025 (DOI: \url{https://doi.org/10.54499/UID/PRR/50021/2025}) and by Blockchain.PT – Decentralize Portugal with Blockchain Agenda, (Project no 51), WP 1: Agriculture and Agri-food, Call no 02/C05-i01.01/2022, funded by the Portuguese Recovery and Resilience Program (PRR), The Portuguese Republic and The European Union (EU) under the framework of Next Generation EU Program.

\bibliographystyle{unsrtnat}
\bibliography{paper} 

\end{document}